\documentclass[10pt]{article}    

\usepackage{amsmath,amsthm,amsfonts,amscd}
\usepackage{eucal}      
\usepackage{verbatim}       
\usepackage{makeidx}        
\usepackage{psfig}          
\usepackage{epsfig}             
\usepackage{citesort}           %
\usepackage{url}        
\usepackage{braket}
\usepackage{amsmath}
\begin{document}
\title{Infinite Component Relativistic Wave
Equations\footnote{Presented at The Dirac Centenary Conference,
Baylor University, Waco, TX  October 1, 2002.}}
\author{Eyal Gilboa\footnote{egilboa@physics.utexas.edu} \\
The University of Texas at Austin}
\date{September 30, 2002}
\maketitle

\begin{abstract}
We construct an infinite component relativistic wave equation
which is a linear first order differential equation identical in
form to a Dirac like equation, describing composite fields
possessing multiple spin and energy states.  The main motivation
for such a construction is to give a description of Hadronic
fields moving along their so called Regge Trajectories, however
this may be generalized to other composite fields.  In order to
construct the equation so that it may accommodate physical states
the concept of Spin Frames is introduced, and it is found that
such an equation may propagate physical fields whose spin states
differ by two units of angular momentum namely $\Delta J=2$.  The
solution for the free field case is given by boosting a rest frame
spinor with the infinite dimensional Lorentz Boost which are
constructed as well.  Finally we discuss the relevance of the
groups GL(4R), and GL(3,1,R) and their appearance with regards to
the wave equation at hand.
\end{abstract}

\section{Introduction}\label{sec4a}
\index{Infinite wave equations}

The construction of relativistic wave equations has been of great
interest to physicists, in particular in the last century where
relativistic quantum mechanics has evolved to the now celebrated
quantum field theory.  Besides the complete relativistic
description of a free filed, the latter has invoked a need for
such equations, so that mechanisms such as free field perturbation
theory will have some degree of success.  In this context much
work has been invested on studying free field equations describing
particles possessing a particular spin
\cite{Dirac1,Bargman,schwinger}(those have been mostly of fields
carrying spin 0, 1, 2, $\frac{1}{2}$, and even $\frac{3}{2}$),
however not as much has focused on relativistic wave equations
describing fields possessing multiple states of spin\footnote{Some
of these constructions were initially made by
Majorana~\cite{Majorana}, and Dirac~\cite{Dirac2}.} and rest mass
(bound states), or so called composite fields. A particular case
in point which we will utilize repeatedly as a fitting case for
such a construction of relativistic wave equations, and in fact
will provide a major motivation for our work is an Hadronic field.
A Hadron is a composite entity which can be excited along a
particular Regge trajectory and therefore may poses different
states both in energy, and in angular momentum. It follows that
the physical motivation for constructing such an equation is to
have a collective description of interactions among such fields
taking into account the various energy and angular momentum states
such fields may acquire.  This description would be an effective
treatment of Hadronic fields whose fundamental interactions are
that of constituent quarks and gluons, those however would not be
specified in such a treatment, but would be inferred from the fact
that excited states of the composite fields correspond to various
interactions of constituent fields.  It is therefore appropriate
that we limit our discussion to kinematic considerations with
respect to such an infinite component wave equation.

To demonstrate further the physical motivation for such an
equation in the context of Hadron physics it would be instructive
to examine the decomposition of physical currents of Hadronic
fields. As an example, a spin $\frac{1}{2}$ field may be used to
demonstrate one such current decomposition \index{Current
decomposition}.

According to the Gordon identity \cite{Peskin} \index{Current
decomposition!Gordon Decomposition} the current for a spin
$\frac{1}{2}$ field is given (in momentum space) by:

\begin{eqnarray}
j^{\mu}(p,p')= \bra{p'}\frac{p'^{\mu}+
p^{\mu}}{2m}+\frac{i\sigma^{\mu\nu}q_{\nu}}{2m}\ket{p}\label{dirac1}
\end{eqnarray}
where $q_{\nu}=p'_{\nu}-p_{\nu}$, and $\sigma^{\mu\nu}$ comprise
the six Lorentz generators. \\

This decomposition is of course based on the existence of a linear
wave equation,  or what is know as the Dirac Equation. What is
revelling about this decomposition is that the current is split
into two parts; one describing the linear motion of the current
given by the first term on the right hand side of~(\ref{dirac1}),
the second describing the spinning part of the current given by
the second term on the right.  The latter also gives us the
magnetic moment of the electron.

It is evident that such a decomposition will describe a particle
of a definite spin (in this case it is $\frac{1}{2}$), and indeed
by the use of the Bargmann Wigner equations~\cite{Bargman} such a
decomposition should be attained for a field of any spin. However,
for a particle acquiring different spin states as it interacts
such a decomposition would have to yield another term to its
current, namely that which has an expectation value between
different angular momentum states, indicating that such a particle
may acquire additional higher multipole moments.  Indeed it was
shown in~\cite{Eyal,Ne'emanchromo} that Hadrons may interact
through their quadrupole moments \index{Quadrupole} produced by a
digluon intermediate state which leads to Hadron structure
deformation. This would be a hint that an additional term
to~(\ref{dirac1}) would have to be added, and would be
proportional to:
\begin{eqnarray}
\bigtriangleup j^{\mu}= \bra{p'j'}c_1\tau_1^{\mu\nu}l_{1\nu}+
c_2\tau_2^{\mu\nu\rho}l_{2\nu}l_{3\rho}+...+
c_n\tau_n^{\mu\nu...\rho_n}l_{2\nu}l_{3\rho}...l_{n\rho_n}\ket{pj}\label{dirac2}
\end{eqnarray}
where the terms $\tau^{\mu\nu...\rho_n}_n$  are tensors associated
with multipole moments\index{multipole} that a particle may
posses, and the vectors $l_{n\rho_n}$ are four vectors comprised
of linear combinations of the vectors $p$ and $p'$.

The first tensor in~({\ref{dirac2}) is a symmetric `shear' tensor
describing quadrupolar pulsations~\cite{Ne'eman2} providing
excitations between angular momentum states that differ by
$\bigtriangleup J=2$ along a given Regge trajectory \index{Regge
trajectory}. Higher terms in~(\ref{dirac2}) will produce
transitions of $\bigtriangleup J=3,4....n$.

In a sense this formalism would be an attempt to give a
description of an entire Regge trajectory, and therefore its
consistency is dependent on the presence of strong interactions.
After all a Regge (Hadronic) field is considered a bound state
very much like a Hydrogen atom whose excitations are dependent on
electromagnetic interactions.  For the present wave equation
(which carries multiple states of angular momentum) to meet its
ultimate goal in describing such excitations, one has to assume as
a priori that any interactions of interest must require that
Hadrons appear as asymptotic states.

\section{A First Construction of the Equation}
\label{sec4b}

\subsection{Preliminary considerations}

In this section we will attempt to construct the infinite
component wave equation more or less in a `naive' approach which
will initially render the field in question un-physical since it
will fail to satisfy a causal condition.  This initial effort
however will not be in vain since it does point the way on how to
overcome such a problem by utilizing the same formalism with some
modifications.

The current~(\ref{dirac2}) suggests that an infinite component
wave equation describing a field possessing multiple spin states
could be a first order linear differential equation similar to
that of the Dirac equation.  In fact the main motivation for such
a construction is derived from Dirac's inspiration for searching
relativistic first order linear equations to describe interaction
of particles which possess spin. We therefore conjecture on the
form of this equation namely:

\begin{equation}
\left(i\partial_{\alpha}^{\mu}X_{\mu}-M_{\alpha}\right)\psi({x_{\alpha}})=0.\label{we1}
\end{equation}

The terms $X^{\mu}$ whose construction will be given below, are
infinite dimensional matrices describing transitions between
different spin states, and are analogous to the gamma matrices
appearing in the Dirac equation. The index $\mu$ is a Lorentz
index, while $\alpha$ is an infinite (spin) frame\index{Infinite
spin frame} index for which we reserve the discussion on its
meaning for later.\footnote{The first letters in the Greek
alphabet will be reserved for frame indices.}.  The term
$M_{\alpha}$ is a real diagonal matrix whose entrees correspond to
the masses of each (spin) frame.

Because we insist that $(M_{\alpha})_{jj}$ correspond to the
masses of the Regge field as it acquires different spin states we
are actually saying that the mass shell condition
$p^2_{\alpha}\psi_j=m^2_{\alpha~j}\psi_j$ is satisfied for each
spin component of the field $\psi$ \index{mass shell condition}.
In fact one of the major driving conditions for building the
representations of $X^{\mu}$ will be this mass shell condition,
however it should be noted that for the wave equation~(\ref{we1})
to be Lorentz invariant such a condition is not necessary.

A famous case in point is the infinite component wave equation
constructed by Majorana\index{Majorana}~ \cite{Majorana} from
which two important features are worth while examining. First, in
his treatment the matrix $M$ is considered to be a constant (and
thus does not carry any spin frame index), which leads to a
particle mass spectrum of the form $m=\frac{M}{j+1}$. This does
not correspond to the observed spectrum of Hadrons, never the less
it is a unique and a remarkable result.  This result also implies
that in the rest frame the operator $p^0$ does not correspond to
$m$, and therefore it follows that Majorana does not enforce the
relation $p^2-m^2=0$.  The second feature of this equation is that
the matrix $X^0$ is positive definite, meaning that in the Fourier
transform of the field $\psi$, negative frequencies are excluded,
implying that quantization of the field $\psi$ would be a bit
awkward since this field would not correspond to neither Fermi nor
Bose Einstein statistics \footnote{I would like to thank E.C.G
Sudarshan for clarifying to me this point.}.

Such features will be avoided from the following treatment by
first introducing an infinite spin frame.  This means that for
each spin state $p^0_{\alpha}$ is different, and if the latter is
to be interpreted as the mass of the particle in the rest frame
then it also must follow that $p^2_{\alpha}=m^2_{\alpha}$. This
means that each spin state has a different mass forcing
$M_{\alpha}$ not to be a constant.

The physical motivation for introducing frames was explained by
Ne'eman \cite{Ne'emaneq} when considering relativistic excitations
of fields which carry multiple angular momentum states. The
analogy is made with that of Einstein's gravity in a local frame
on a 4-manyfold with a defined local metric using tetrads.  In
this formalism any general transformation under the covariant
group (diffeomorphisms) can describe a transformation from one
local frame to another in a gravitational field.  In our case,
excitations of Hadronic fields really correspond to structural
deformations of these `extended' objects. Since the problem of
describing relativistic extended objects (in four dimensions) is
yet to be formulated,  spin is one of the local properties that
may be used to describe such excitations. Similar to the situation
in gravity, one can describe the transformation of a four manyfold
of an extended object from one geometrical state (frame) to
another by describing the spin states associated with each
geometrical state with the use of some symmetry group.  Since an
Hadronic field in theory can be excited to an infinite amount of
geometrical states (frames) through structural deformations, where
each geometrical state corresponds to a particular spin state
(frame); hence the term infinite spin frames.

It should be stressed that an infinite (spin) has no relation and
should not to be confused with a Lorentz frame.  The indices
$\mu$, and $\alpha$ represent two distinct coordinates; one
representing a space time coordinate, the other a spin coordinate.
The relation between $\alpha$ the spin frame, and $j$ the spin of
the particle will be given in section~(\ref{sec4d}).

\subsection{The representations of
$X^{\mu}$}\index{Representations of $X^{\mu}$}

Since the matrices $X^{\mu}$ should comprise a four vector (a
statement which has yet to be proven) it should follow
that:\index{Four vector}
\begin{equation}
\left[S^{\mu\nu},X^{\lambda}\right]=i\eta^{\nu\lambda}X^{\mu}-i\eta^{\mu\lambda}X^{\nu}\label{lorentz}
\end{equation}
In light of this the following spherical basis\index{Four
vector!sperical basis} can be defined:

\begin{eqnarray}
X^+&=&X^1+iX^2\nonumber\\
X^-&=&X^1-iX^2\nonumber\\
X^3&=&X^3\nonumber\\
X^0&=&X^0. \label{add1}
\end{eqnarray}

This together with~(\ref{lorentz}), the general representations of
$X^{\mu}$ can be written by the use of the Wigner Eckart theorem.
One particular representation which will be convenient for the
development of the subject has been given by Weyl~\cite{Weyl} for
the spherical basis~(\ref{add1}), and is given by:
\begin{eqnarray}
\bra{\alpha,j,m}X^+\ket{\alpha,j-1,m-1}&=&-c^-(\alpha,j)\sqrt{(j+m)(j+m-1)}\nonumber\\
\bra{\alpha,j,m-1}X^-\ket{\alpha,j-1,m}&=&c^-(\alpha,j)\sqrt{(j-m+1)(j-m)}\nonumber\\
\bra{\alpha,j,m}X^3\ket{\alpha,j-1,m}&=&c^-(\alpha,j)\sqrt{(j+m)(j-m)}\nonumber\\
\bra{\alpha,j,m}X^+\ket{\alpha,j,m-1}&=&c(\alpha,j)\sqrt{(j+m)(j-m+1)}\nonumber\\
\bra{\alpha,j,m-1}X^-\ket{\alpha,j,m}&=&c(\alpha,j)\sqrt{(j-m+1)(j+m)}\nonumber\\
\bra{\alpha,j,m}X^3\ket{\alpha,j,m}&=&c(j)m\nonumber\\
\bra{\alpha,j,m}X^+\ket{\alpha,j+1,m-1}&=&c^+(\alpha,j)\sqrt{(j-m+1)(j-m+2)}\nonumber\\
\bra{\alpha,j,m-1}X^-\ket{\alpha,j+1,m}&=&-c^+(\alpha,j)\sqrt{(j+m)(j+m+1)}\nonumber\\
\bra{\alpha,j,m}X^3\ket{\alpha,j-1,m}&=&c^+(\alpha,j)\sqrt{(j+m+1)(j-m+1)}\nonumber\\
\bra{\alpha,j,m}X^0\ket{\alpha,j,m}&=&c^0(\alpha,j)\nonumber\\
\label{wig}
\end{eqnarray}

\noindent The term general representations with regards to
~(\ref{add1})) is used since the coefficients $c^+(\alpha,j)$,
$c^-(\alpha,j)$, $c(\alpha,j)$, and $c^0(\alpha,j)$ have yet to be
determined.  As will be shown these representations are
proportional to the infinite dimensional Lorentz
boosts~\cite{Chandra}, just as in the case of the Dirac equation
where the gamma matrices are a result of boosts acting on a field.

The coefficient $c(\alpha,j)$ furnishes a finite representation
of~(\ref{we1}) and therefore is of no use to us in the current
development \footnote{it will be shown below that in fact an
inclusion of such terms is inconsistent with certain conditions
the wave equation must fulfill}.  To obtain the coefficients
$c^+(\alpha,j)$, $c^-(\alpha,j)$, and $c^0(\alpha,j)$ we enforce
the following condition:
\begin{equation}
\left(\partial_{\alpha}^2+m^2_{\alpha}\right)\psi=0.
\label{masssell}
\end{equation}
\footnote{There are more conditions related to Lorentz invariance
(to be discussed) that the mass matrix must fulfill in order to be
called a mass matrix, though~(\ref{masssell}) will suffice to
obtain the coefficients $c^+(\alpha,j)$, $c^-(\alpha,j)$, and
$c^0(\alpha,j)$.} To arrive at such a condition one can use
Dirac's trick by operating on~(\ref{we1}) from the left with its
conjugate for a particular frame and spin to give:

\begin{eqnarray}
\left[\left(-i\partial_{\alpha}^{\nu}X_{\nu}-M_{\alpha}\right)\left(i\partial_{\alpha}^{\mu}X_{\mu}-M_{\alpha}\right)
\right]_{j,j}\psi({x_{\alpha}})=\nonumber\\
\left[\frac{1}{2}\partial^{\nu}_{\alpha}\partial^{\mu}_{\alpha}\left\{X_{\nu}X_{\mu}\right\}+m^2_{\alpha}\right]_{j,j}
\psi({x_{\alpha}})=0,\label{we2}
\end{eqnarray}

\noindent and it appears that in order to arrive at a mass shell
condition the $X^{\mu}$'s have to fulfill the familiar statement
for each spin frame namely,
\begin{equation}
\bra{\alpha,j,m}\left\{X_{\nu}X_{\mu}\right\}\ket{\alpha',j,m}=2\eta_{\mu\nu}\delta_{\alpha\alpha'}.\label{dirac3}
\end{equation}
It should be emphasized that unlike the gamma matrices appearing
in the Dirac equation, the statement
$\left\{X_{\nu}X_{\mu}\right\}=2\eta_{\mu\nu}$ is not true in
general, but should only apply when this expression is taken
between equal values of spin.

Keeping this in mind we note that when $\mu\neq\nu$, it is easy to
see that the condition~(\ref{dirac3}) is satisfied.  First, for
the case $\mu=0,\nu=i$, or vice a versa the term
\begin{displaymath}
\bra{\alpha,j,m}X^0X^i\ket{\alpha,j,m}=0.
\end{displaymath}
This is because according to~(\ref{lorentz}), $X^0$ is spherical
tensor of rank zero, while $X^i$ is a spherical tensor of rank
one.  This term therefore can only have expectation values between
$j$ and $j\pm 1$.  Second, for the case where both $\mu$ and $\nu$
are space indices one can decompose their bilinear as follows:
\begin{equation}
X^iX^j=\frac{1}{3}\delta^{ij}D_3
+\frac{1}{2}\left[X^iX^j\right]+\frac{1}{2}\left(\left\{X^iX^j\right\}-\frac{2}{3}D_3\right)\label{bilin},
\end{equation}
where $D_3=X^iX_i$.

The nine bilinears according to the decomposition of~(\ref{bilin})
have split into one spherical scalar, three spherical vectors, and
five spherical tensors of rank two containing the symmetric
combination in~(\ref{dirac3}).  When $i\neq j$ the latter
symmetric combination can only have expectation values between $j$
and $j\pm 2$, and condition~(\ref{dirac3}) is satisfied for this
case.

For the case $\mu=\nu=0$ one gets from~(\ref{dirac3}) that
$c^0(\alpha,j)=\pm 1$. For the case $\mu=\nu=i$ the situation is
more problematic for which the representations of $X^{\mu}$ will
have to be modified. To illustrate how naive our approach has been
so far in dealing with this construction, and the problem that
arises it is sufficient to evaluate the term $X^{i}X_{i}$. Using
the representations~(\ref{wig}), the following is obtained
\footnote{The terms proportional to $c(\alpha,j)$ have been
included in this step to justify the earlier assumption made on
their irrelevance.}:
\begin{eqnarray}
-D_3 &=& -(j(j+1)+m)c^2(\alpha,j)-c^-(\alpha,j)c^+(\alpha,j-1)(j+m)(2j-1)\nonumber\\
 &+& c^+(\alpha,j)c^-(\alpha,j+1)(j-m+1)(2j+3)-m\nonumber\\
 \label{ex1}
\end{eqnarray}

\noindent This expression should not depend neither on $j$, nor on
$m$, in fact the expression above should equal $-X^{i}X_{i}=3$.
Further, $c^0(\alpha,j)$, $c^+(\alpha,j)$, $c^+(\alpha,j)$, and
$c(\alpha,j)$ are all independent of $m$, therefore
from~(\ref{ex1}) the following two conditions are obtained:
\begin{eqnarray}
1+c^2(\alpha,j)&=&c^+(\alpha,j)c^-(\alpha,j+1)(2j+3)\nonumber\\
&-&c^-(\alpha,j)c^+(\alpha,j-1)(2j-1),\label{ex2a}
\end{eqnarray}
and
\begin{eqnarray}
-D_3&=&-j(j+1)c^2(\alpha,j)-c^+(\alpha,j)c^-(\alpha,j+1)(j+1)(2j+3)\nonumber\\
&-&c^-(\alpha,j)c^+(\alpha,j-1)j(2j-1)\label{ex2b}.
\end{eqnarray}
These relations enable to eliminate the term $c(\alpha,j)$
completely, which does not have any baring on the representation
of $X^i$, so it might as well be set equal to zero.  Hence the
only possible way to satisfy the condition $D_3=-3$ would be to
set

\begin{equation}
c^+(\alpha,j)c^-(\alpha,j+1)=-c^-(\alpha,j)c^+(\alpha,j-1)=\frac{1}{2j+1}\label{ex3}
\end{equation}
This implies however that:
\begin{equation}
c^+(\alpha,j+1)c^-(\alpha,j+2)=c^-(\alpha,j)c^+(\alpha,j-1)\label{ex4}
\end{equation}
which says that $c^+(\alpha,j)$, and $c^-(\alpha,j)$ are both
independent of $j$ contradicting~(\ref{ex3}).  Although an
impossible result, statements~(\ref{ex3}, \ref{ex4}) do suggest
that the form $c^+(\alpha,j)\sim c^-(\alpha,j-1) \sim
\frac{1}{\sqrt{2j+1}}$ up to some phase is the right prescription
if there was another set of coefficients requiring more freedom
with respect to how one writes the representations for $X^i$. Thus
there should exist $c^+_1(\alpha,j_1,j_2)$,
$c^+_2(\alpha,j_1,j_2)$, $c^-_1(\alpha,j_1,j_2)$,
$c^-_2(\alpha,j_1,j_2)$, for each $X^i$, and $c^0_1(\alpha,j_1),
c^0_2(\alpha,j_2)$ for each $X^0$ with ($|j_1-j_2|=1$) . Thus
expression~(\ref{ex3}) would read:

\begin{eqnarray}
c^+_1(\alpha,j_1)c^-_2(\alpha,j_2)=-c^-_1(\alpha,j_1)c^+_2(\alpha,j_2)=\frac{\eta}{2j_1+1}\nonumber\\
c^+_1(\alpha,j_2)c^-_2(\alpha,j_1)=-c^-_1(\alpha,j_2)c^+_2(\alpha,j_1)=\frac{\eta}{2j_2+1},\label{ex5}
\end{eqnarray}
where $\eta$ is some phase, and the contradiction of the form
of~(\ref{ex4}) is avoided.  Although these additions are motivated
by the above arguments they have not been fully justified, and it
seems a bit blurry at this point how such conditions may arise
from the representations given by~(\ref{wig}).  To put these on a
more firm footing we proceed at this stage to the next section to
better formulate the representations of $X^{\mu}$.

\section{A Second Construction of the Equation}
\label{sec4c}
\subsection{Defining the odd and even representations}

As indicated in the last section it seems that there should be two
sets of matrices $X^{\mu}$ in order to fulfill~(\ref{dirac3}).  It
will be shown that having two of these matrices corresponds to a
kind of chirality that exist in the Dirac case, where the gamma
matrices in the chiral representation mix left and right
components of the spinor field.  The correspondence between the
current wave equation and the Dirac equation is that the two sets
of $X^{\mu}$ will act as the sigma matrices appearing in the gamma
matrices of the chiral representation, performing the mixing
mentioned above.\footnote{There may well be other infinite
representations not necessarily of this type that may fulfill
conditions that have been mentioned (in
particular~(\ref{dirac3})), though I am not aware of any.}.

We proceed with this construction by first defining the projection
operator:\index{projection operator}

\begin{equation}
\bra{\alpha,j,m'}P_n\ket{\alpha',j',m'}=\delta_{\alpha\alpha'}\delta_{j,j'}\delta_{m,m'},\label{proj}
\end{equation}
with $n=j-j_{min}$, and where $j_{min}$ is the lowest value of $j$
for a particular representation.  This matrix operator has
$(2j+1)\times(2j+1)$ dimensions with $m=j,j-1...-j$ which forms a
subspace of the infinite dimensional identity, hence
\begin{displaymath}
I=\sum_{n=0}^{\infty}P_n.\label{defofproj}
\end{displaymath}

Using~(\ref{defofproj}) it is possible to define the odd, and even
projection operators respectively $P_{2n\pm1}$, and $P_{2n}$ which
have the property of splitting the Hilbert space $\ket{jm}$ into
two separate spaces which we will call `odd' and `even'
respectively. These are given by:
\begin{eqnarray}
\ket{j_{o},m}=P_o\ket{j,m}=\sum_{n}^{\infty}P_{2n+1}\ket{j,m}\nonumber\\
\ket{j_{e},m}=P_e\ket{j,m}=\sum_{n}^{\infty}P_{2n}\ket{j,m}\label{evod}.
\end{eqnarray}
where $|j_o-j_e|=1$.

\noindent Thus for example for a space with half integral spins
\begin{eqnarray}
j_e=\frac{1}{2}, \frac{5}{2}, \frac{9}{2}...\nonumber\\
j_o=\frac{3}{2}, \frac{7}{2}, \frac{11}{2}...
\end{eqnarray}

Now define the following operators:

\begin{eqnarray}
L^i&=&\sum_n^{\infty}~P_{2n+1}X^iP_{2n}+P_{2n}X^iP_{2n+1}\label{lsproj}\\
N^i&=&\sum_n^{\infty}~P_{2n-1}X^iP_{2n}+P_{2n}X^i_{2n-1}\label{nsproj}\\
L^0&=&\sum_n^{\infty}P_{2n+1}X^0P_{2n+1}\label{ltproj}\\
N^0&=&\sum_n^{\infty}~P_{2n}X^0P_{2n}\label{ntproj},
\end{eqnarray}\\
where it is understood that for the lowest term in~(\ref{nsproj})
$P_{-1}=0$.

\noindent From this splitting~(\ref{evod}) it is apparent that
with respect to rotations the operators $L^i, N^i$, and $L^0, N^0$
are spherical tensors of rank one and zero respectively, and by
the methods of~({\ref{wig}) their representations are given
by\index{Representation of $X^{\mu}$}:

\begin{eqnarray}
\left. \begin{array}{ccc}
\bra{\alpha,j_o,m}L^+\ket{\alpha,j_e,m-1}&=&-l^-(\alpha,j_o)\sqrt{(j_o+m)(j_o+m-1)}\nonumber\\
\bra{\alpha,j_o,m-1}L^-\ket{\alpha,j_e,m}&=&l^-(\alpha,j_o)\sqrt{(j_o-m+1)(j_o-m)}\nonumber\\
\bra{\alpha,j_o,m}L^3\ket{\alpha,j_e,m}&=&l^-(\alpha,j_o)\sqrt{(j_o+m)(j_o-m)}\nonumber\\
\bra{\alpha,j_e,m}L^+\ket{\alpha,j_o,m-1}&=&l^+(\alpha,j_e)\sqrt{(j_o-m-1)(j_o-m)}\nonumber\\
\bra{\alpha,j_e,m-1}L^-\ket{\alpha,j_o,m}&=&-l^+(\alpha,j_e)\sqrt{(j_o+m-1)(j_o+m)}\nonumber\\
\bra{\alpha,j_e,m}L^3\ket{\alpha,j_o,m}&=&l^+(\alpha,j_e)\sqrt{(j_o+m)(j_o-m)}\nonumber\\
 \bra{\alpha,j_e,m}L^0\ket{\alpha,j_o,m}&=&l^0(\alpha,j_o)\end{array}\right\}
j_o-j_e=1,\nonumber\\
\label{lrep}
\end{eqnarray}

\noindent and
\begin{eqnarray}
\left. \begin{array}{ccc}
\bra{\alpha,j_e,m}N^+\ket{\alpha,j_o,m-1}&=&-n^-(\alpha,j_e)\sqrt{(j_e+m+1)(j_e+m)}\nonumber\\
\bra{\alpha,j_e,m-1}N^-\ket{\alpha,j_o,m}&=&n^-(\alpha,j_e)\sqrt{(j_e-m+1)(j_e-m)}\nonumber\\
\bra{\alpha,j_e,m}N^3\ket{\alpha,j_o,m}&=&n^-(\alpha,j_e)\sqrt{(j_e+m)(j_e-m)}\nonumber\\
\bra{\alpha,j_o,m}N^+\ket{\alpha,j_e,m-1}&=&n^+(\alpha,j_o)\sqrt{(j_e-m)(j_e-m-1)}\nonumber\\
\bra{\alpha,j_o,m-1}N^-\ket{\alpha,j_e,m}&=&-n^+(\alpha,j_o)\sqrt{(j_e+m)(j_e+m+1)}\nonumber\\
\bra{\alpha,j_o,m}N^3\ket{\alpha,j_e,m}&=&n^+(\alpha,j_o)\sqrt{(j_e+m)(j_e-m)}\nonumber\\
\bra{\alpha,j_o,m}N^0\ket{\alpha,j_e,m}&=&n^0(\alpha,j_e)
\end{array}\right\}j_e-j_o=1.\nonumber\\
\label{nrep}
\end{eqnarray}\\
\noindent The terms $l(\alpha,j), n(\alpha,j)=0$ by the same
arguments of the previous section.

It follows from the representations~(\ref{lrep}), and~(\ref{nrep})
that the bilinears of the operators $L^{\mu}$, $N^{\mu}$ have
expectation values only between the following states:

\begin{eqnarray*}
\bra{j,m}L^{\mu}L^{\nu}\ket{j'=j,m'=m,m\pm2},&\hspace{0.5cm}&\bra{j}N^{\mu}N^{\nu}\ket{j'=j,m'=m,m\pm2},\\
\bra{j}L^iN^j\ket{j'=j\pm2,m'=m,m\pm2},&\hspace{0.5cm}&\bra{j}N^iL^j\ket{j'=j\pm2,m'=m,m\pm2}.
\label{prop}
\end{eqnarray*}

With these representations the matrices $X^{\mu}$ can be
constructed by adding a two dimensional index to their labels
denoted by $a,b=1,2$ to give $X^{\mu}_{ab}$, which are defined as
follows:

\begin{eqnarray}
X^{i}_{12}&=&L^{i}+N^{i}\nonumber\\
X^{i}_{21}&=& L^{i}-N^{i}\nonumber\\
X^i_{11}&=&X^i_{22}=0.\nonumber\\
 \label{chiralrep}
\end{eqnarray}
and
\begin{eqnarray}
X^0_{11}&=&X^0_{22}=\frac{1}{2}(L^0+N^0)\nonumber\\
X^0_{12}&=&X^0_{21}=0.
\end{eqnarray}

\noindent The above representations could also be written more
compactly as:

\begin{eqnarray}
\bra{\alpha,j,m}X^{\mu}\ket{\alpha'j',m'}=\bra{\alpha,j,m}\left(\begin{array}{cc}
        X^0_{11} & X^{i}_{12}\\ X^{i}_{21} & X^0_{22}
        \end{array}\right)\ket{\alpha',j',m'}
\end{eqnarray}

In defining the multiplication with regards to the indices $a,b$
we introduce a raising and lowering metric\index{metric
$\epsilon_{ab}$} $\epsilon_{ab}=\epsilon^{ab}$, where
$\epsilon_{11}=\epsilon_{22}$, $\epsilon_{12}=\epsilon_{21}=0$,
and whose representation is given by
\begin{equation}
\epsilon_{ab}=\varepsilon_{\alpha,\alpha'j,j',m,m'}\delta_{ab},\label{metr2}
\end{equation}
where
\begin{eqnarray}
\varepsilon_{\alpha\alpha'j,j',m,m'}= \left\{ \begin{array}{cc}
\delta_{\alpha,\alpha'}\delta_{j,j'}\delta_{m,m'}&j=j_o\\-\delta_{\alpha,\alpha'}\delta_{j,j'}\delta_{m,m'}&j=j_e
\end{array}\right.\label{metr1}
\end{eqnarray}

The metric $\epsilon_{ab}$ has the following properties:
\begin{eqnarray}
\{\epsilon_{ab},X^i\}=0\nonumber\\
\left[\epsilon_{ab},X^0\right]=0,\label{coan}
\end{eqnarray}
and behaves like the $4\times4$ matrix $\gamma^0$ that appears in
the Dirac theory.

Using the metric~(\ref{metr1}) the following bilinear can be
defined:

\begin{equation}
Q^{\mu\nu}_{ab}=(X^{\mu}X^{\nu})_{ab}=X^{\mu}_{ac}\epsilon^{cc'}X^{\nu}_{c'b}.\label{metr3}
\end{equation}





The condition~(\ref{dirac3}) can now be implemented, and due to
~(\ref{prop}) only terms proportional to $(L^i)^2$ and $(N^i)^2$
need to be considered when evaluating the following:

\begin{eqnarray}
(D_4)_{ab}&=&\bra{\alpha,j,m}(X^{\mu}X_{\mu})_{ab}\ket{\alpha,j,m}=\bra{\alpha,j,m}X^{\mu
~2}_{1}X_{\mu~21}
\ket{\alpha,j,m}\nonumber\\
&=&\bra{\alpha,j,m}X^{\mu 1}_{2}X_{\mu~12}\ket{\alpha,j,m}.\label{inv}\\
\end{eqnarray}

In order to determine the coefficients $l^{\pm}(\alpha,j),
n^{\pm}(\alpha,j), l^0(\alpha,j)$, and $n^0(\alpha,j)$ it would be
sufficient to evaluate $(X^1)^2$, (any other of the space
components will do as well), and $(X^0)^2$. Using the
representation~(\ref{lrep}), and~(\ref{nrep}) together
with~(\ref{inv}) it follows that

\begin{eqnarray}
\bra{\alpha,j,m}(L^1+N^1)(L^1-N^1)\ket{\alpha,j,m}&=&\nonumber\\
&=&\frac{1}{2}\Big[l^-(\alpha,j)l^+(\alpha,j-1)(j^2-j+m^2)\nonumber\\
&-&n^+(\alpha,j)n^-(\alpha,j+1)(j^2+3j+m^2+2)\Big].\nonumber\\
\label{con1}
\end{eqnarray}

\noindent Because $l^{\pm}(\alpha,j)$, and $n^{\pm}(\alpha,j)$,
are independent of $m$ one immediate consequence of~(\ref{con1})
is
\begin{equation}
n^+(\alpha,j)n^-(\alpha,j+1)=l^-(\alpha,j)l^+(\alpha,j-1).\label{con2}
\end{equation}

\noindent Now~(\ref{con1}) reads
\begin{equation}
\bra{\alpha,j,m}(L^1+N^1)(L^1-N^1)\ket{\alpha,j,m}=-l^-(\alpha,j)l^+(\alpha,j-1)(2j+1).\label{con3}
\end{equation}

\noindent By choosing

\begin{eqnarray}
l^-(\alpha,j)=l^+(\alpha,j-1)=\frac{\eta}{\sqrt{(2j+1)}}\nonumber\\
n^+(\alpha,j)=
n^-(\alpha,j+1)=\frac{\eta}{\sqrt{(2j+1)}}\label{con4}
\end{eqnarray}

\noindent with $\eta$ being either a pure complex number, or a
pure real number with a modulus of unity, the
condition~(\ref{dirac3}) can be satisfied for all the space
components of $X^{\mu}$ up to a sign.  The choice $\eta=i$ will
make $\bra{j,m}(X^i)^2\ket{j,m}=1$ (up to a sign), but more
importantly would mean that the matrices $X^i$ are anti-Hermitian
which would be consistent if $\psi$ in~(\ref{we1}) is to be
interpreted as a field.

For the time component $X^0$, the condition~(\ref{dirac3}), and
the condition that in the rest frame $p_0^{\alpha}=m^{\alpha}$
implies that
\begin{equation}
n^0(\alpha,j)=l^0(\alpha,j)=1\label{time}
\end{equation}
\noindent With this condition~(\ref{dirac3}) reads as follows:
\begin{equation}
\bra{\alpha,j,m}\left\{X_{\nu}X_{\mu}\right\}\ket{\alpha,j,m}=
2\bigtriangleup _{\mu\nu},\label{dirac4}
\end{equation}
where
\begin{eqnarray}
\bigtriangleup_{\mu\nu}=\left\{ \begin{array}{cc}
 \eta_{\mu\nu}&j=j_o\\
-\delta_{\mu\nu}&j=j_e\end{array}\right.\nonumber\\
\label{met0}
\end{eqnarray}

\noindent This peculiarity has a physical consequence and in fact
is necessary for any physical interpretation as will be discussed
in the next section.  Although the choice of sign in~({\ref{time})
is arbitrary, other conventions will not change the outcome
~({\ref{met0}), namely that the metric changes from a Euclidian
metric to a Minkowskian metric for states that differ by
$\bigtriangleup J=1$ in~(\ref{met0}).

\subsection{Lorentz Generators}
\index{Lorentz generators}

At this stage it would be useful to look at commutation relations
of the type $\left[X^{\mu},X^{\nu}\right]_{ab}$, where one
contraction with respect to the indices $a, b$ is obtained, so as
to project a (1,1) component with respect to these indices. To get
a feel what such terms correspond to it would be instructive to
evaluate as an example the following commutator: \index{Lorentz
Generators}
\begin{eqnarray}
\left[X^1,X^2\right]_{11}&=& X^1_{12}\epsilon_{22}X^2_{21}-X^2_{12}\epsilon_{22}X^1_{21}\nonumber\\
&=&\epsilon_{22}\left( X^2_{12}X^1_{21}-  X^1_{12}X^2_{21}\right)
\label{comu1}
\end{eqnarray}

\noindent Implementing the definitions~(\ref{chiralrep}), and a
bit of algebra we get:
\begin{eqnarray}
\left[X^1,X^2\right]_{11}&=&-i\frac{\epsilon_{22}}{2}\Big(N^+N^- - N^-N^+ L^-L^- -L^+L^-\nonumber\\
&+& L^+N^- - L^-N^+ -N^+L^- + N^-L^+ \Big)\nonumber\\
&=&-i\frac{\epsilon_{22}}{2}\Big(\left[N^+,N^-\right]+\left[L^-,L^+\right]+\left\{L^+,N^-\right\}\nonumber\\
&-&\left\{L^-,N^+\right\}\Big).\label{comu2}
\end{eqnarray}

\noindent Using the representations~(\ref{nrep}), (\ref{lrep}),
and the properties~(\ref{prop}) the four expressions
in~(\ref{comu2}) can be evaluated for their non-vanishing matrix
elements which are given by (we drop the labels `odd', and `even'
from $j$):
\begin{eqnarray}
\bra{j,m}\left[N^+,N^-\right]\ket{j,m}&=&2m\left(\frac{2j+3}{2j+1}\right)\nonumber\\
\bra{j,m}\left[L^-,L^+\right]\ket{j,m}&=&-2m\left(\frac{-2j+1}{2j+1}\right)\nonumber\\
\bra{j,m}L^+,N^-\ket{j-2,m}&=&\bra{j,m}L^-N^+\ket{j-2,m}\nonumber\\
&=&\sqrt{\frac{(j+m)(j+m-1)(j-m)(j-m-1)}{(2j+1)(2j-3)}}\nonumber\\
\bra{j,m}N^+L^-\ket{j+2,m}&=&\bra{j,m}N^-L^+\ket{j+2,m}\nonumber\\
&=&\sqrt{\frac{(j+m+1)(j+m+2)(j-m+1)(j-m+2)}{(2j+1)(2j+3)}}.\nonumber\\
 \label{comu3}
\end{eqnarray}

\noindent The first two terms in~(\ref{comu2}) add, while the last
two exactly cancel, and we are left with
\begin{eqnarray}
\frac{i}{2}\left[X^1,X^2\right]_{11}=\frac{i}{2}\left[X^1,X^2\right]_{22}\nonumber\\
=\bra{j,m}J_3\ket{j,m}=m,\label{comu4}
\end{eqnarray}

\noindent which is nothing more than the representation of the
3-generator of angular momentum.  Upon evaluating
\begin{eqnarray}
\frac{i}{2}\Big(\left[X^2,X^3\right]_{ab}+i\left[X^1,X^3\right]_{ab}\Big)
&=&\bra{j,m}J^+\ket{j,m-1}\nonumber\\
&=&\sqrt{(j+m)(j-m+1)}\nonumber\\
\frac{i}{2}\Big(\left[X^2,X^3\right]_{ab}-i\left[X^1,X^3\right]_{ab}\Big)
&=&\bra{j,m}J^-\ket{j,m+1}\nonumber\\
&=&\sqrt{(j+m+1)(j-m)}\nonumber\\
 \label{comu5}
\end{eqnarray}

\noindent it is evident that the terms
$\frac{i}{2}\left[X^i,X^j\right]_{ab}$ furnish the finite
spinorial representations of the group of rotation $SU(2)$.

By the same method, and upon the use of
representation~(\ref{lrep}, \ref{nrep}) the following operators
are evaluated
\begin{equation}
K^i=S^{i0}_{ab}=\frac{i}{2}\left[X^i,X^0\right]_{ab}.\label{comuex}
\end{equation}

\noindent Evaluating the commutation relations
\begin{equation}
\left[K^i,K^j\right]=-i\epsilon_{ijk}J^k\hspace{0.4cm}\left[K^i,J^j\right]=i\epsilon_{ijk}K^k
\label{comuex2}
\end{equation}
it follows that the operators $K^i$ furnish the representations of
the boost generators of the Lorentz group.  It should be noted
that the representation of $\epsilon_{ab}$~({\ref{metr2})
establishes the consistency of these commutation relations, and
provides the motivation for its definition.

\section{Physical States}
\label{sec4d}

\subsection{`Even' and `odd' trajectories and spin frames}

In this section we would like to investigate the physical
consequences of equation~(\ref{we1}) in light of the
representations and properties of $X^{\mu}$ discussed in the
previous sections.

Equation~(\ref{we1}) can be written in component form which
manifestly resembles a chiral component of the Dirac equation in
its chiral representation, (while providing an entire different
physical description) and is given in the form:

\begin{eqnarray}
\left(\begin{array}{cc} \partial_{\alpha}^0X^0
-M_{aa~\alpha}&-{\mathbf{\nabla}}_{\alpha}\cdot{\mathbf{
 X}}_{ab}\\
-{\mathbf{\nabla}}_{\alpha}\cdot{\mathbf{
 X}}_{ba}&\partial_{\alpha}^0X^0-M_{bb~\alpha}
 \end{array}\right)_{j j'}\left(\begin{array}{c}
 \psi^a(x_{\alpha})\\
\psi^b(x_{\alpha})\end{array}\right)_{j'}=0. \label{we3}
 \end{eqnarray}
where it is understood that $a, b=1,2$, and that upper and lower
indices are connected via the metric~(\ref{metr2}), with the term
$M_{ab~\alpha}=
\delta_{ab}\delta_{j,j'}m_{\alpha}$.\\

 As stated in section~(\ref{sec4b}) the mass shell condition
 should be satisfied for each frame, meaning each
 state of a definite spin should satisfy $p^2_{\alpha}=m^2_{\alpha}$.
 This is done by effecting the operation~(\ref{we2})
 which gives the second order equation

\begin{eqnarray}
\left(\begin{array}{cc}
(\partial^0)^2_{\alpha}\epsilon_{aa}-\nabla^2_{\alpha}I_{aa}+\epsilon_{aa}M^2_{\alpha}&0\\
0&(\partial^0)^2_{\alpha}\epsilon_{bb}-\nabla^2_{\alpha}+\epsilon_{bb}M^2_{\alpha}\end{array}\right)_{jj}
\left(\begin{array}{c}
 \psi^a(x_{\alpha})\\
\psi^b(x_{\alpha})\end{array}\right)_{j}=0,\nonumber\\
 \label{we4}
\end{eqnarray}
where $I_{aa}$ is just the identity matrix.\\

We have attempted to achieve a Klein-Gordon equation however the
result in~(\ref{we4}) is not quite such an equation if the term
$M_{\alpha}$ is to be interpreted as the mass of the particle in a
particular spin state.  The source of what seems to be a problem
is the alternating signs of the terms
$(\partial^0)^2_{\alpha}\epsilon_{ab}$ and $\epsilon_{ab}M^2$  as
one goes from $j_o$ to $j_e$.  By the definition of
$\epsilon_{ab}$~(\ref{metr2}) it is obvious that for $j=j_o$ one
has the equation

\begin{equation}
\left(\partial^2_{\alpha}+M^2_{\alpha}\right)_{j_oj_0}\psi_{j_o}=0,
\label{confl1}
\end{equation}

\noindent while for $j=j_e$ Eq.~(\ref{we4}) reads

\begin{equation}
\left((\partial^0)^2_{\alpha}+\nabla^2_{\alpha}+M^2_{\alpha}\right)_{j_ej_e}\psi_{j_e}=0.
\label{confl2}
\end{equation}

\noindent So it is evident that states that carry $j=j_o$ do
satisfy the Klein-Gordon equation while those that carry $j=j_e$
do not.  It should be stressed that had a different convention
been chosen for the metric, $j_e$ would satisfy~(\ref{confl1}),
while $j_o$ would satisfy~(\ref{confl2}).  As an example, in the
former case $\psi_{j_o}$ would be a trajectory of states carrying
half integral spin $\frac{3}{2}, \frac{7}{2}, \frac{11}{2}.....$
and does satisfy the Klein-Gordon equation.  On the other hand
$\psi_{j_e}$ is a trajectory of states carrying the half integral
spins $\frac{1}{2}, \frac{5}{2}, \frac{9}{2}.....$, and which does
not satisfy the Klein-Gordon equation. The natural question to ask
now is why aren't these two trajectories on the same physical
footing?

To resolve this quandary one has to look more carefully at the
Lorentz transformations furnished by the representations of
$X^{\mu}$, particularly the boosts.  Under an infitesimal Lorentz
boost $\psi$ transforms as\index{Lorentz boost}

\begin{eqnarray}
\psi'(x^{\alpha})&=&1+iS^{i0}_{ab}\hspace{0.1cm}\phi_{i0}\hspace{0.2cm}\psi(x^{\alpha})\nonumber\\
&=&1-\frac{1}{2}\left[X^{i},X^{0}\right]_{ab}\hspace{0.05cm}\phi_{i0}\hspace{0.2cm}\psi(x^{\alpha}),\nonumber\\
\label{confl3}
\end{eqnarray}

\noindent and it is evident from the representations~(\ref{lrep},
\ref{nrep}) that the above transformation is non-unitary since the
Lorentz boost are anti-Hermitian.  Furthermore, according to these
representations a Lorentz boost takes the field from a state $j$
to a state $j\pm1$.  From a physical stand point there is
absolutely no reason why one observer may measure one value of
spin in one frame, and a second observer may measure a different
value in another (boosted) frame. The boost being represented
non-unitarily is exactly the right prescription since these
correspond to non-physical transitions.  What seems as a conflict
in~(\ref{confl2}) is simply a statement that states that
correspond to $j=j_e$ (keeping in mind our initial convention of
$\epsilon_{ab}$) are not physical states, and therefore do not
have a rest frame description nor a mass.  As a result
equation~(\ref{we3}) applies to fields whose states differ in two
units of angular momentum.

The concept of infinite spin frames \index{Spin frames} which has
been used repeatedly so far becomes more clear. It is
phenomenologically well known that a Hadron can be excited to
different spin states when interacting with an external field, due
to its structure (comprised of gluons and quarks).  In
calculations of physical processes such as amplitudes and cross
sections, one should take into account all possible states that
the particle can be found in, which provides the main motivation
for the wave equation at hand.  A Regge trajectory can be defined
as a sum on an infinite ensemble of free states, or frames (in
$J/m^2$ space). In other words if a particle is free, it is said
to be in a specific spin frame as far as its spin and rest mass
are concerned. Applying a boost to a particle similar to the one
in~(\ref{confl3}) alters its Lorentzian (kinetic) frame, however
it does not alter its spin frame.  The fact that the boost
in~(\ref{confl3}) may change the field's angular momentum
component by $\bigtriangleup J=\pm1$ has no physical consequence
what so ever since the transition to such a state is done
non-unitarily, and therefore it not observed.  Thus a particular
spin frame which is described by a Regge field will span angular
momentum values $J-1<J<J+1$.  If a Regge field is measured to be
in a physical state with some spin $J$ then by the argument
presented above its $J-1, J+1$ components cannot correspond to any
physical observables, and therefore don't have a rest frame
description in the equation~(\ref{we1}).

The field describing an entire Regge trajectory as a sum of spin
frames thus may be written as:
\begin{equation}
\Psi^a(x)=\sum^{\infty}_{\alpha=0}\sum_j\psi^a_j(x_{\alpha})
\label{regtra1}
\end{equation}
\noindent where $j$ is either $j_o$, or $j_e$, and the sum on
their values is given by:\\

\begin{equation}
2\alpha+(j_o)_{min}-1 \leq j_o \leq
2\alpha+(j_o)_{min}+1,\label{regtra2}
\end{equation}
for and odd trajectory, while for an even trajectory

\begin{equation}
2\alpha-(j_e)_{min} \leq j_o \leq
2\alpha-(j_e)_{min}+2.\label{regtra3}
\end{equation}

\noindent In the latter case when $\alpha =0$ the left hand side
of~(\ref{regtra3}) should be taken as zero.

\subsection{Left and right boosts and free particle solutions}
\index{Free particle solutions}

We now proceed to obtain free particle solutions to the wave
equation~({\ref{we1}) by boosting a rest frame spinor. Before
performing such a boost it would be worth while to further
investigate the properties of the Lorentz generators given
by~(\ref{comu8}). One of these is given by:

\begin{equation}
S^{i0}_{12}=\frac{i}{2}(X^i_{12}\epsilon_{22}X^0_{22}-X^0_{11}\epsilon_{11}X^i_{12}).\label{left1}
\end{equation}
\noindent From the properties of the metric $\epsilon_{ab}$ given
by~({\ref{coan}) it is evident that~(\ref{left1}) furnishes two
representations of a boost which could be obtained by
anti-commuting the metric $\epsilon_{ab}$ either to the left, or
to the right; thus yielding two representations that differ by an
over all minus sign (the same goes for $S^{i0}_{21}$).  Due to
this ambiguity it is appropriate to define left and right boosts
given by: \index{Left and right boosts}
\begin{eqnarray}
(S^{i0}_{ab})_{LL}&=&\epsilon^{cc'}X^{i}_{ac}X^0_{c'b}\nonumber\\
(S^{i0}_{ab})_{RR}&=&X^{i}_{ac}X^0_{c'b} \epsilon^{cc'},\nonumber\\
\label{leri1}
\end{eqnarray}

\noindent The representations~(\ref{leri1}) suggests that the wave
equation~({\ref{we1}) supports left and right fields which
transform differently under Lorentz boosts according to
\begin{eqnarray}
\psi(x^{\alpha})_{aR}&=&1+i(S^{i0}_{ab})_{RR}\phi^{\alpha}_i\xi\nonumber\\
\psi(x^{\alpha})_{aL}&=&1+i(S^{i0}_{ab})_{LL}\phi^{\alpha}_i\xi,\nonumber\\
\label{boost1}
\end{eqnarray}
where $\xi$ is a rest frame spinor with $(2j+1)\times 1$
dimensions.

\noindent Splitting the above two components we find the
following:
\begin{eqnarray}
(\psi_1)_R=1-\frac{1}{2}X^i_{12}\phi_i\xi\hspace{0.4cm}(\psi_2)_R=1-\frac{1}{2}X^i_{21}\phi_i\xi,\nonumber\\
\label{boost2}
\end{eqnarray}
and
\begin{eqnarray}
(\psi_2)_L=1+\frac{1}{2}X^i_{12}\phi_i\xi\hspace{0.4cm}(\psi_2)_L=1+\frac{1}{2}X^i_{21}\phi_i\xi,\nonumber\\
\label{boost3}
\end{eqnarray}

\noindent  Because of the left/right splitting the wave equation
~(\ref{we1}) doubles in dimension with respect to indices $ab$
which is analogous to the Dirac case in the chiral representation.
The matrices $X^{\mu}$ now take on the form\footnote{With this
definition $X^i_{21}$ is different from its initial definition by
a minus sign.}:
\begin{equation}
X^i_{abRL}=-X^i_{abLR} \hspace{0.3cm} X^0_{aaRL}=X^0_{aaLR},
\label{lrex}
\end{equation}
or in a left/right tensor form
\begin{eqnarray}
X^i=\left(\begin{array}{cc}
 0&X_{RL}^i\\-X_{LR}^i&0\end{array}\right)\hspace{0.3cm}
 X^0=\left(\begin{array}{cc}0&X^0_{RL}\\X^0_{LR}&0\end{array}\right)
\label{mer1}
\end{eqnarray}
with
\begin{eqnarray}
X^i_{RL}=\left(\begin{array}{cc}
 0&X^i_{12}\\X^i_{21}&0
 \end{array}\right)\hspace{0.3cm}X^0_{RL}=\left(\begin{array}{cc}1&0\\0&1
 \end{array}\right)
\label{mer2}
\end{eqnarray}

 \noindent The representation~(\ref{mer1}) assume that
multiplication of any two $X^{\mu}_{ab}$ is done between left and
right components which induces a left/right metric $g_{LR}$. Thus
the four dimensional space $abLR$ is very similar in nature to
that of the four dimensional space that exist in Dirac case of the
chiral representation.

Equation~(\ref{we1}) in the rest frame with both left and right
components takes a finite dimensional form and can be written as
the following (in momentum space):

\begin{eqnarray}
\left(\begin{array}{cc}
 -M_{\alpha}&p^0_{\alpha}\\
 p^0_{\alpha}&-M_{\alpha}
 \end{array}\right)_{jj}\left(\begin{array}{c}
\psi_R\\
\psi_L \end{array}\right)_j=0,
 \label{free1}
\end{eqnarray}

\noindent where $\psi$ has $2j+1$ components, and the matrix above
$2(2j+1) \times 2(2j+1)$ dimensions (for a spin $\frac{1}{2}$ the
above is identical to the rest frame \index{Rest frame} Dirac
equation).  The vanishing determinant of this matrix produces
$2(2j+1)$ eigenvalues; $2j+1$ with a value of $M_{\alpha}$, and
$2j+1$ with a value of $-M_{\alpha}$, implying that the
equation~(\ref{we1}) admits negative particle solutions

To obtain a solution to the free particle wave equation in a
moving frame we can apply a Lorentz boosts to the rest frame
spinor, keeping in mind that such boosts must not take the field
out of its spin frame.  Effectively what needs to be done is to
decompose the full Lorentz transformation with respect to each
spin frame, and to project those spherical tensors which don't
take the particle out of its spin frame.  To understand this
better we write the Lorentz transformations as the following:

\begin{eqnarray}
U_{Lorentz}&=&e^{iS^{\mu\nu}\omega_{\mu\nu}}=1+iS^{\mu\nu}\omega
_{\mu\nu}+\frac{1}{2}(iS^{\mu\nu}\omega _{\mu\nu})^2\nonumber\\
&+&\frac{1}{6}(iS^{\mu\nu}\omega_{\mu\nu})^3 +.... \label{boost4}
\end{eqnarray}

\noindent With respect to the different powers of $S^{\mu\nu}$ it
is evident that each power can be decomposed into its irreducible
spherical tensors.  For example the first term in~(\ref{boost4})
is a scalar, the second is a vector, the third is comprised of
rank two tensor and a scalar, the fourth is comprised of a vector
and a rank three tensor, and so on.  For each term in the series
above that contains a spherical tensor of rank $n$, a field with a
spin component of $J$ will be taken by the action of each of these
terms to a state of $J \pm n$, each being independent from the
other. If the wave equation~(\ref{we1}) is to describes a free
particle in a specific state, and this equation at most connects
field components that differ in angular momentum by
$\bigtriangleup J= 0,\pm1$, with a physical spin frame spanning
components of $J-1<J<J+1$, it follows that the only relevant
tensors (as far as a physical boost is concerned) are those that
could carry the field components from $J$ to $ J\pm 0,1$, namely
those tensors which are scalars or vectors.  If higher order terms
are to be considered for each frame, for example tensors of rank
two from the third term in~(\ref{boost4}), then the free wave
equation would describe transitions between physical frames.
Physically this wouldn't be consistent since a wave equation of a
non-interacting particle under a Lorentz transformation must
retain its description of the same particle of the same spin.  In
effect this means that one can choose a subset of solutions to the
wave equation which correspond to the physical solutions while
neglecting those that correspond to non-physical solutions.  From
this picture it stems that a particular Lorentz transformation can
describe a particular boost for an infinite amount of spin frames,
however each one will be disjoint from the
other as required by physical constraints. \\

 In order to extract from the boosts~(\ref{boost4}) the
appropriate components fitting each spin, the Lorentz boosts
generators (in our chiral representation~(\ref{leri1})) can be
written as the following:
\begin{equation}
{\mathbf{X}}_{abRL}\rightarrow
(P^{\alpha}{\mathbf{X}}P^{\alpha})_{abRL}, \label{boost4a}
\end{equation}
where $P^{\alpha}$ is the projection operator which is defined in
a similar way to that in~(\ref{proj}), and $\alpha$ determines the
range of $J$ given by~(\ref{regtra2}).  This does not alter the
representation of the boosts what so ever since the upper and
lower values of the representations~(\ref{lrep}, \ref{nrep}) for
any particular $J$ are within the limits of~(\ref{regtra2}).

\noindent Now apply this boost as in ~(\ref{boost4}) to a rest
frame spinor:
\begin{eqnarray}
\psi(p^{\alpha})_{a~j}&=&\left(e^{iS^{\mu\nu}\omega_{\mu\nu}}\right)_{physical}\xi\nonumber\\
&=&\sum_b \sum_j \Big(1-(P^{\alpha}{\mathbf{X}}P^{\alpha})_{ab~jj'}\cdot \left(\frac{{\bf{\phi}}}{2}\right)\nonumber\\
&+&\frac{1}{2}(P^{\alpha}{\mathbf{X}}P^{\alpha})^2_{ab~jj'}\cdot\left(\frac{{\mathbf{\phi}}}{2}\right)^2+.....\Big)\xi_{j_o}.\nonumber\\
 \label{boost5}
\end{eqnarray}
\noindent The second term just gives the original boost back while
the third term gives:
\begin{eqnarray}
\frac{1}{2}\sum_{jm}\sum_{b'}P^{\alpha}_{aajjmm}(X^i)_{ab'jj''mm''}P^{\alpha}_{b'b'j''j''m''m''}
(X^j)_{b'bj''j'm''m'}P^{\alpha}_{bbj'j'm'm'}
\frac{\phi_i}{2}\frac{\phi_j}{2}\nonumber\\
 =
 \delta_{ab}\delta_{jj'}\delta_{m'm'}\frac{1}{2}\left(\frac{\phi}{2}\right)^2\nonumber\\
\label{boost6}
\end{eqnarray}
where it is assumed that $j'$ belongs to a physical state, meaning
the one which actually propagates and for which
$(\epsilon_{ab})_{j'j~m'm}=1$. Because of~(\ref{boost6}) the
fourth term in the series gives:
\begin{eqnarray}
\frac{1}{3!}(P^{\alpha}{\mathbf{X}}P^{\alpha})^3_{ab~jj'}=\frac{1}{3!}{\mathbf{X}}
,\nonumber\\
\label{boost7}
\end{eqnarray}
and when summing on all terms in the series the right handed
boosted field is given by:
\begin{eqnarray}
\psi(p^{\alpha})_{aR~j}&=&\left(e^{iS^{\mu\nu}\omega_{\mu\nu}}\right)_{physical}\xi\nonumber\\
&=&\left({\rm{cosh(\phi)}}-
{\mathbf{X}}_{RL}\cdot{\mathbf{n}}~{\rm{sinh(\phi)}}\right)\xi.
\label{boost8}
\end{eqnarray}
Putting the expressions for the rapidity and the explicit
expression for ${\mathbf{X}}_{RL}$ together with a summation on
$\alpha$, (\ref{boost8}) can be written as:
\begin{eqnarray}
\left(\begin{array}{c} \Psi_1(p)\\
\Psi_2(p)\end{array}\right)_R=\sum_{\alpha}\frac{1}{\sqrt{2m^{\alpha}(E^{\alpha}+m^{\alpha})}}
\left(\begin{array}{cc}
E^{\alpha}+m^{\alpha}&-{\mathbf{p}}\cdot{\mathbf{X_{12}}}\\
-{\mathbf{p}}\cdot{\mathbf{X_{21}}}&E^{\alpha}+m^{\alpha}\end{array}\right)\left(\begin{array}{c}\xi_1\\\xi_2
\end{array}\right),\nonumber\\
\label{boost9}
\end{eqnarray}
while the left handed field is given by
\begin{eqnarray}
\left(\begin{array}{c} \Psi_1(p)\\
\Psi_2(p)\end{array}\right)_L=\sum_{\alpha}\frac{1}{\sqrt{2m^{\alpha}(E^{\alpha}+m^{\alpha})}}
\left(\begin{array}{cc}
E^{\alpha}+m^{\alpha}&{\mathbf{p}}\cdot{\mathbf{X_{12}}}\\
{\mathbf{p}}\cdot{\mathbf{X_{21}}}&E^{\alpha}+m^{\alpha}\end{array}\right)\left(\begin{array}{c}\xi_1\\\xi_2
\end{array}\right),\nonumber\\
\label{boost10}
\end{eqnarray}

The above representation establishes the fact that under a
physical Lorentz transformation the following is true
\begin{equation}
\left(U^{-1}_{Lorentz}X^{\mu}_{ab}U_{Lorentz}\right)_{Physical}=\Lambda^{\mu}_{\nu}X^{\nu}_{ab},\label{inv}
\end{equation}
where $\Lambda^{\mu}_{\nu}$ are the familiar four dimensional
Lorentz transformation.   This establishes that the wave
equation~(\ref{we1}) is Lorentz invariant.

To complete the construction of the wave equation~(\ref{we1}) we
need to address the question of what values should the mass matrix
$M^{\alpha}_{ab}$ take for those entries which correspond to
non-physical states.  For example lets assume that particles
corresponding to the trajectory $j_o$ are the physical states then
according to~(\ref{boost8}, \ref{boost9}) it is clear that
\begin{equation}
(M^{\alpha}_{11})_{j_oj_o}=(M^{\alpha}_{22})_{j_oj_o}=m^{\alpha}_{j_o}.\label{mass1}
\end{equation}

\noindent Inserting~(\ref{boost8}, \ref{boost9}) into the wave
equation~(\ref{we1}) we also find
\begin{equation}
(M^{\alpha}_{11})_{j_o\pm1j_o\pm1}=(M^{\alpha}_{22})_{j_o\pm1j_o\pm1}=m^{\alpha}_{j_o}.\label{mass2}
\end{equation}
\noindent These relations are also consistent with~(\ref{inv}),
meaning the mass matrix in equation ({\ref{we1}) transforms as a
scalar under a Lorentz transformation.

\section{The  GL(3,1R) and GL(4,R) Groups}
\label{sec4e}
\subsection{ GL(3,1)/GL(4R) splitting}

In the introduction it was proposed that the non-locality of
Hadrons may manifest itself in local parameters such as spin. If
equation~(\ref{we1}) describes a collection of states acquired by
a Regge field then it is reasonable to assume that this equation
could give rise to currents discussed in~(\ref{dirac2}) when
interactions are concerned (for which no treatment will be given
in the present work).  Due to this and the fact that the
representation of the Lorentz generators are proportional to the
commutators of $X^{\mu}X^{\nu}$ suggest that
these bi-linears may have further properties worth exploring.\\

In section~(\ref{sec4b}) the tensorial decomposition of the
bi-linear $X^iX^j$ was performed, and was given by: \footnote{ Due
to the left/right representations of the Lorentz transformation it
will be convenient in what follows to contract such bi-linears
with the metric $\epsilon_{ab}$ from either the left or the right.
Also since $\epsilon_{ab}$ commutes with such terms the labels $a,
b$ are suppressed, and it is understood that all multiplications
are done with this metric between left/right, and right/left
components.}:
\begin{equation}
X^iX^j=\frac{1}{3}\delta^{ij}D_3
+\frac{1}{2}\left[X^iX^j\right]+\frac{1}{2}\left(\left\{X^iX^j\right\}-\frac{2}{3}D_3\right)\label{bilin2},
\end{equation}
where $D_3=X^iX_i$.

The decomposition~(\ref{bilin2}) splits the bilinear $iX^iX^j$
into three irreducible spherical tensors.  The first is a
symmetric tensor $D_3\eta^{\mu\nu}$ which has one independent
term, the second an anti-symmetric traceless tensor $S^{ij}$ which
has three terms, and the last is a traceless symmetric tensor
$T^{ij}$ which has five terms.  Given this, and the
representations~(\ref{lrep}, \ref{nrep}), it follows that the
first term transforms as a spherical tensor of rank zero, the
second term transforms as a spherical tensor of rank one, and the
third as a spherical tensor of rank two, all under
rotations, making nine terms all together.\\

\noindent Now define the following irreducible spherical tensors:
\begin{eqnarray}
T^0_0&=&D\nonumber\\
J^k&=&\epsilon_{kij}S^{ij}\nonumber\\
J^+&=&J^1+iJ^2  \hspace{1cm} J^-=J^{1}-iJ^{2}\nonumber\\
T^2_0&=&\frac{1}{\sqrt{6}}\left(T_{11}+T_{22}-2T_{33}\right)\nonumber\\
T^2_{+1}&=&T_{13}+iT_{23}\nonumber\\
T^2_{+1}&=&T_{13}+iT_{23}\nonumber\\
T^{2}_{+2}&=& T_{11}-T_{22}+2iT_{22}\nonumber\\
T^{2}_{-2}&=& T_{11}-T_{22}-2iT_{22}\nonumber\\
\label{sphert1}
\end{eqnarray}
 By the Wigner Eckart theorem the following commutation
relations follow:
\begin{eqnarray}
\big[J^3,T^0\big]&=&0 \nonumber\\
\big[J^3,T^j_m \big] &=& mT^j_m \nonumber\\
\big[J^{\pm},T^1_m \big] &=& \sqrt{(1 \mp m)(1\pm m+1)}T^1_{m\pm1}\nonumber\\
\big[J^{\pm},T^2_m\big] &=& \sqrt{(2 \mp m)(2\pm m+1)}T^2_{m\pm1}.\nonumber\\
\label{comu6}
\end{eqnarray}
\noindent The last three of these could also be put in the form:
\begin{eqnarray}
\big[J^i,T^{jk}\big]=i\epsilon^{ijl}T^{lk}+i\epsilon^{ikl}T^{jl}.\nonumber\\
\label{comu7}
\end{eqnarray}

\noindent Since the angular momentum operator is given
by~(\ref{comu5}), then~(\ref{comu7}) can also be written as:

\begin{eqnarray}
\big[S^{ij},S^{kl}\big]&=&-i\eta^{ik}S^{jl}+i \eta^{il}S^{jk}\nonumber\\
&+&i\eta^{jk}S^{il}-i\eta^{jl}S^{ik}\nonumber\\
\big[S^{ij},T^{kl}\big]&=&-i\eta^{ik}T^{jl}-i\eta^{il}T^{jk}\nonumber\\
&+&i\eta^{jk}T^{il}-i\eta^{jl}T^{ik}\nonumber\\
\big[S^{ij},D_3\big]&=&0\nonumber\\
\big[T^{ij},D_3\big]&=&0\nonumber\\
\label{comu8}
\end{eqnarray}

\noindent  These commutation relations are satisfied if the
operators:
\begin{equation}
Q^{ij}=\frac{i}{2}X^{i}X^{j}
\end{equation}
\noindent(note that this definition of $Q^{ij}$ is different from
that of~(\ref{metr2}) by a factor of $\frac{1}{2}$) satisfy the
commutation relations: \index{GL($3R$) commutation relations}

\begin{equation}
\left[Q^{ij},Q^{kl}\right]=i\eta^{jk}Q^{il}-i\eta^{il}Q^{kj},
\label{gl3}
\end{equation}

\noindent which are the commutation relations of the group
$GL(3R)$, and which make $Q^{ij}$ a $GL(3R)$ group element.
Omitting the operator $D_3$, the irreducible spherical tensors
~(\ref{sphert1}) furnish the representation of the group $SL(3R)$.
This group has been shown by Ne'eman \cite{Ne'eman2} to give the
correct Regge excitations of Hadrons with respect to angular
momentum, and its emergence should not come as a big surprise.
Extended objects unlike point like particles posses properties
that are indicative of their structure; one such example is the
particle's spin altered by some deformation of its structure.  The
operators $T^{ij}$ are the shear tensors that describe such
deformations, and lead to excitations of angular momentum with
$\bigtriangleup J=2$.  Indeed such excitations characterize a
Regge trajectory \cite{Chew}, where resonances of
Hadrons are observed to have these selection rules. \\

 We can attempt to do the same kind of analysis above while
incorporating the operators $Q^{i0}$ $Q^{00}$, however
from~(\ref{lrep}, \ref{nrep}) it is apparent that these are
already irreducible.   Furthermore there seems to be a peculiarity
in the way these are defined with respect to the metric
$\epsilon_{ab}$.  Unlike the operators $Q^{ij}$, multiplication
for left or right by $\epsilon_{ab}$ doesn't affect the
representations of these operators since $\epsilon_{ab}$
anti-commutes with $X^i$.  On the other hand for the operators
$Q^{i0}$ we have:
\begin{equation}
(Q^{i0})_{abLL}=\frac{i}{2}\epsilon_{cc'}X^i_{ac}X^0_{c'b}=-(Q^{i0})_{abRR}=
\frac{i}{2}X^i_{ac}X^0_{c'b}\epsilon_{cc'}\label{gl41}
\end{equation}
\noindent For the case where multiplication is done from the left
one finds the commutation
relations:\index{GL($4R$)/GL($3,1R$)commutation relations}
\begin{equation}
\left[Q^{\mu\nu},Q^{\lambda\sigma}\right]=i\delta^{\nu\lambda}Q^{\mu\sigma}-i\delta^{\mu\sigma}
Q^{\lambda\nu}, \label{gl42}
\end{equation}
where $-\delta^{\mu\nu}=(-1, -1, -1, -1)$ is the four dimensional
(negative) Euclidean metric, which would indicate that
$(Q^{\mu\nu})$ is a $GL(4R)$ group element.  However for the right
hand multiplication we get the same commutation relation as
in~({\ref{gl42}) but with $-\delta^{\mu\nu}$ replaced with
$\eta^{\mu\nu}=(+1, -1, -1, -1 )$ the Minkowskian metric,
indicting that $Q^{\mu\nu}$ is a $GL(3,1)$ group element.  The
source of this splitting can be understood by observing that
\begin{equation}
\bra{\alpha,j,m}\{X^{\mu}X^{\nu}\}\ket{\alpha,jm}=\bigtriangleup^{\mu\nu}
\label{gl43}
\end{equation}
where $\bigtriangleup^{\mu\nu}_j$ is $\eta^{\mu\nu}$ if $J$
corresponds to an angular momentum of physical state, or
$\delta^{\mu\nu}$ if $J$ corresponds to angular momentum of a
non-physical state.  Again as an example choosing $J=j_o$ to be a
physical state described by the wave equation~({\ref{we1}) would
mean it's free particle description is governed by the subgroup of
$GL(3,1)$, namely $SO(3,1)$ with $\eta^{\mu\nu}$ being its
invariant metric. Upon the action of a boost whose representations
are made from bi-linears of $X^{\mu}$, a physical state is carried
to a non-physical state in a non-unitary fashion and its `free'
description is now described by the subgroup of $GL(4R)$, namely
$SO(4)$ with a negative metric; hence this state cannot correspond
to any physical observable.  Finally  if a Wick rotation of the
form
\begin{equation}
X^0\rightarrow iX^0 \hspace{0.4cm} X^i\rightarrow X^i \label{wick}
\end{equation}
is performed then the left and right multiplications with respect
to the metric $\epsilon_{ab}$ are interchanged, and so do the
roles of $\eta^{\mu\nu}$, and $-\delta^{\mu\nu}$.  This should be
expected since the symmetry groups $GL(3,1R)$, and $GL(4R)$ are
separated by such a rotation.

\subsection{Establishing $X^{\mu}$ as Lorentz vector}

Finally we would like to show that the commutation
relations~(\ref{gl42}) (with $\eta^{\mu\nu}$ being the metric) are
consistent with the statement that $X^{\mu}$ is indeed a Lorentz
vector, which simply means the following \cite{Peskin}:

\begin{eqnarray}
\left[S^{\mu\nu},X^{\lambda}\right]=i\eta^{\nu\lambda}X^{\mu}-i\eta^{\mu\lambda}X^{\nu}.
\label{lorentz1}
\end{eqnarray}
Although this statement has already been shown for each spin frame
in~(\ref{inv}) by a decomposition of a full Lorentz transformation
with respect to each frame, it would be more desirable to show it
in a more general fashion.

The above relation could be verified by using the
representations~(\ref{lrep}, \ref{nrep}), which cloud turn out to
be a lengthy process due to the latter being of the infinite
dimensional type, involving Clebsch-Gordan coefficients, never the
less it can be done. Fortunately there is a more efficient way to
establish~(\ref{lorentz1}) which utilizes
~(\ref{gl42}).\\

We start by writing the left hand side of ~(\ref{gl42}) as the
following:
\begin{eqnarray}
\left[Q^{\mu\nu},Q^{\lambda\sigma}\right]&=&\frac{i}{2}\big[Q^{\mu\nu},X^{\lambda}\big]X^{\sigma}\nonumber\\
&+&\frac{i}{2}X^{\lambda}\big[Q^{\mu\nu}X^{\sigma}\big]\nonumber\\
\label{lorentz2}
\end{eqnarray}\\
Without any loss of generality the commutators on the right hand
side of~(\ref{lorentz2}) may be written in the following way:

\begin{eqnarray}
\big[Q^{\mu\nu},X^{\lambda}\big]=i\eta^{\nu\lambda}X^{\mu}-i\eta^{\mu\lambda}X^{\nu}
+\Theta^{\mu\nu\lambda},\nonumber\\
\label{lorentz3}
\end{eqnarray}
\noindent with an analogous expression for
$\big[Q^{\mu\nu}X^{\sigma}\big]$, and we note that the tensor
$\Theta^{\mu\nu\lambda}$ is a reducible rank three tensor with no
apparent symmetries.\\

\noindent Putting~(\ref{lorentz3}) back into~(\ref{lorentz2}), and
using~(\ref{gl42}) the following is obtained:

\begin{eqnarray}
\frac{i}{2}\Theta^{\mu\nu\lambda}X^{\sigma}+\frac{i}{2}X^{\lambda}\Theta^{\mu\nu\sigma}
&=&\eta^{\mu\lambda}Q^{\nu\sigma}-\eta^{\nu\sigma}Q^{\lambda\mu}\nonumber\\
&=&-i\left[Q^{\nu\mu},Q^{\lambda\sigma}\right]\nonumber\\
\label{lorentz5}
\end{eqnarray}

\noindent This suggests that the tensors $\Theta^{\mu\nu\lambda}$,
 can be written as the following:

\begin{eqnarray}
\Theta^{\mu\nu\lambda}&=&\frac{1}{2}\Big(X^{\nu}X^{\mu}X^{\lambda}
-X^{\lambda}X^{\nu}X^{\mu}+\Pi^{\mu\nu\lambda}\Big),\nonumber\\
 \label{lorentz6}
\end{eqnarray}
again with a similar expression for $\Theta^{\mu\nu\sigma}$, and
$\Pi^{\mu\nu\lambda}$ being a general a tensor with no apparent
special properties.  Putting these expressions on the left hand
side of~(\ref{lorentz5}), and upon comparing to the right hand
side while writing $Q^{\mu\nu}$ in terms of $X^{\mu}X^{\nu}$, it
follows that:
\begin{eqnarray}
\Pi^{\mu\nu\lambda}X^{\sigma}+X^{\lambda}\Pi^{\mu\nu\sigma}=0.\label{lorentz7}
\end{eqnarray}

\noindent Interchanging the indices $\lambda$ with $\sigma$, a
similar equation to~(\ref{lorentz7}) is obtained from which it
follows that:

\begin{equation}
\left[\Pi^{\mu\nu\lambda},X^{\sigma}\right]=0.\label{lorentz8}
\end{equation}

\noindent Setting $\lambda=\sigma$, relations~(\ref{lorentz7},
\ref{lorentz8}), and~(\ref{lorentz6}) produce:
\begin{equation}
\Pi^{\mu\nu\lambda}X^{\lambda}=0.\label{lorentz9}
\end{equation}
\noindent From which it follows that
\begin{equation}
\Pi^{\mu\nu\lambda}=0,\label{lorentz10}
\end{equation}
\noindent and
\begin{equation}
\Theta^{\mu\nu\lambda}=\left[X^{\nu}X^{\mu},X^{\lambda}\right]\label{lorentz11}
\end{equation}

\noindent Putting~(\ref{lorentz11}) in~(\ref{lorentz6}) one
obtains:
\begin{equation}
\left[S^{\mu\nu},X^{\lambda}\right]=i\eta^{\nu\lambda}X^{\mu}-i\eta^{\mu\lambda}X^{\nu}.
\label{lorentz12}
\end{equation}
\noindent Hence $X^{\mu}$ is a Lorentz vector.

\section{Concluding Remarks}
In this chapter we have constructed a relativistic infinite
component wave equation describing a field carrying multiple
states of angular momentum.  The physical motivation for such a
construction is to give a collective description for Hadronic
fields which are composite fields exhibiting excitations in both
angular momentum, and rest mass ($J/m^2$) known as excitations
along Regge trajectories.

The treatment above has been purely kinematic, and a fundamental
assumption has been made regarding the compositeness of the field
in questions enabling it to acquire different states of spin.  For
such a relativistic equation to be consistent it was necessary to
introduce the concept of spin frames, a direct result of the
varying mass matrix appearing in~(\ref{we1}).  The physical
motivation for introducing spin frames comes about from the
non-unitarity of the Lorentz transformation constructed from the
infinite dimensional matrices $X^{\mu}$ appearing in~(\ref{we1}).
Under a boost, the field $\psi$ is taken to an angular momentum
state which differs from its original (pre-boosted) state by
$\bigtriangleup J=\pm1$.  Such a transition does not correspond to
an actual physical process, meaning the field under a boost
remains with the same value of spin, or in other words every
observer along any constant moving frame measures the same value
of spin for the field in question.  Thus it was shown that a free
field is characterized by a spin frame which spans spin values
$J-1<J<J+1$, with $J$ being the actual physical spin of the free
field.  As a result of this it was shown that the relativistic
wave equation can support fields whose physical spin states differ
by $\bigtriangleup J=2$.

Due to these kinematic considerations it is apparent that when
interactions are included the field $\psi$ may be excited to
states that differ by $\bigtriangleup J=2$ through the action of
the quadrupolar tensor $T^{\mu\nu}$ which is also constructed form
the matrices $X^{\mu}$.  It therefore seems that structural
(non-local) deformations lead to excitations of spin as should be
expected.  In an interacting theory it is conceivable that
incorporation of fields like $\psi$ with other fields (possibly
gauge fields) may lead to structural information regarding
Hadronic fields through such transitions.  As mentioned in the
introduction of this chapter, having a linear relativistic wave
equation similar to that of the Dirac equation should give rise to
a current decomposition similar to that of the Gordon
identity~(\ref{dirac1}).  Since the minimal physical transition in
spin states according to the wave equation~(\ref{we1}) are in two
units of angular momentum, one should expect that a current
decomposition similar to the latter should yield terms
proportional to the quadrupolar moments of the Hadron.  Of course
a full current decomposition should yield higher multipole
moments.

Although we have been referring to Hadron physics through out this
paper, the construction of the infinite component wave equation is
a general construction that may fit other fields with the same
characteristics.  It is conceivable that a particle such as a
quark, or an electron at some energy (not presently attainable)
would get excited to a higher state of spin and rest mass due to
some internal structure yet to be discovered.


\bibliography{dirac}
\bibliographystyle{plain}
\end{document}